\documentclass[review]{elsarticle}
\usepackage{mdframed}
\usepackage{amsmath,amssymb}
\usepackage{lipsum}
\usepackage{bm}
\usepackage{graphicx}
\usepackage{graphics}
\usepackage{amsmath}
\usepackage{array}
\usepackage[caption=false]{subfig}
\usepackage{xcolor}
\usepackage{subfig}
\usepackage{hyperref}
\usepackage[capitalise]{cleveref}
\usepackage{textcomp}
\usepackage{epstopdf}
\usepackage{cases}
\usepackage{amssymb}
\usepackage{multirow}
\usepackage{siunitx}
\usepackage{cases}
\usepackage{tabularx}
\usepackage[top=1.1in, bottom=1.1in, left=1.0in, right=1.0in]{geometry}
\usepackage[utf8]{inputenc}
\usepackage{tikz}
\usepackage{comment}
\usepackage{lineno,hyperref}
\modulolinenumbers[5]
\usepackage{booktabs}
\renewcommand{\arraystretch}{1.0} 

\begin{document}

\title{Non-equilibrium transport and phonon branch-resolved size effects based on a multi-temperature kinetic model}
\author[add2]{Chuang Zhang}
\ead{zhangc520@hdu.edu.cn}
\author[add4]{Houssem Rezgui}
\ead{houssem.rezgui@inl.int}
\author[add5]{Meng Lian}
\ead{lianmeng@hust.edu.cn}
\author[add2]{Hong Liang\corref{cor1}}
\ead{lianghongstefanie@163.com}
\address[add2]{Department of Physics, School of Sciences, Hangzhou Dianzi University, Hangzhou 310018, China}
\address[add4]{International Iberian Nanotechnology Laboratory (INL),  Braga 4715-330, Portugal}
\address[add5]{School of Physics, Institute for Quantum Science and Engineering and Wuhan National High Magnetic Field Center, Huazhong University of Science and Technology, Wuhan 430074, China}
\cortext[cor1]{Corresponding author}
\date{\today}

\begin{abstract}

A multi-temperature kinetic model is developed for capturing the branch-resolved size effects and electron-phonon coupling process.
Compared with typical macroscopic multi-temperature models, the assumption of diffusive phonon transport is abandoned and replaced by the free migration and scattering of particles.
The phonon branch- and size-dependent effective thermal conductivity is predicted in nanosized graphene as well as the temperature slips near the boundaries.
Compared with other phonon branches, the ZA branch contributes the most to thermal conduction regardless of system sizes.
Furthermore, in nanosized homogeneous graphene with a hotspot at the center, the branch-dependent thermal conductivity increases from the inside to the outside even if the system size is fixed.
The thermal conductivity of ZA branch is even higher than the lattice thermal conductivity when the system size is hundreds of nanometers.

\end{abstract}

\maketitle

\section{Introduction}

Electron-phonon coupling, one of the most fundamental energy exchange methods in nature, is ubiquitous in heat dissipation in electronic devices~\cite{pop_energy_2010,adv_sci_2025_nonequilibrium,HU2025126679,CHEN2024109042}, ultra-fast laser heating or detection~\cite{beardo_hydrodynamic_EP_2025,karna_direct_2023,Sciadv_2019_hotelectron,PhysRevB.77.075133,MTP_review_nuo_2021}, low-dimensional materials~\cite{balandin_superior_2008,graphene_review_conduction,nano_letters_2017_non_equilibrium,Distinguishing_AdvSci_2020,exp_photon_excitation2021} and other fields~\cite{PhysRevLett.68.2834,TTM_EP_AIPX2022}.
For example, electrons absorb most of the energy from photons and further transfer the energy to optical and acoustic phonons through electron-phonon coupling in ultra-fast laser heating experiments~\cite{nano_letters_2017_non_equilibrium,Distinguishing_AdvSci_2020,exp_photon_excitation2021,PhysRevB.93.125432,PhysRevB.98.134309}. 
Electron-phonon coupling process significantly affects the electrical and thermal properties of the materials~\cite{TTM_EP_AIPX2022,Zhu_2024,nano_letters_2017_non_equilibrium}.
Understanding the electron-phonon coupling mechanism and achieving its regulation are beneficial to the design of advanced materials and promotes energy conversion and utilizations~\cite{ChenG05Oxford,pop_energy_2010,balandin_superior_2008,PhysRevLett.130.256901}.

In order to accurately capture the electron-phonon coupling process, one of the most widely used models is the macroscopic two-temperature model~\cite{anisimov1974electron,PhysRevLett.58.1680,PhysRevB.50.15337,QIU19942789,PhysRevB.77.075133,PhysRevB.65.214303,TTM_EP_AIPX2022}, in which the electronic subsystem and phonon subsystem are invoked and a single coefficient is used for representing their interactions.
This model assumes that the thermal equilibrium is arrived for each subsystem and both electron and phonon suffer diffusive transport processes.
However, many experiments have proven that there are obvious non-equilibrium phonon effects for semiconductor materials with weak electron-phonon coupling~\cite{nano_letters_2017_non_equilibrium,Distinguishing_AdvSci_2020,exp_photon_excitation2021}. 
For example, phonon temperature between different branches are different in laser-irradiated single-layer suspended graphene.
It is difficult to accurately characterize branch-resolved energy exchange process with the macroscopic two-temperature diffusive model.

In the past decades, predecessors have made great improvements for better explaining the experimental results~\cite{PhysRevB.93.125432,PhysRevX.6.021003,tong2022multitemperature,PhysRevB.107.L041407}.
Vallabhaneni $et~al$.~\cite{PhysRevB.93.125432} and Lu $et~al$.~\cite{PhysRevB.98.134309} developed a macroscopic multi-temperature model to explain the thermal conduction properties of single-layer suspended graphene in Raman spectroscopy experiments.
Compared with the macroscopic two-temperature model, this multi-temperature model still uses the diffusive transport approximation. 
The main difference is that multiple phonon branch temperatures are introduced, assuming that thermal equilibrium is arrived for each phonon branch, and the interaction coefficients between different phonon branches and electrons are different. 
This model well reflects the different coupling strengths between electrons and optical phonons and acoustic phonons in single-layer suspended graphene or other low-dimensional materials, and has also been adopted by multiple research teams to explain Raman experimental data~\cite{nano_letters_2017_non_equilibrium,Distinguishing_AdvSci_2020,exp_photon_excitation2021}.
Using similar assumptions of diffusive transport and multiple phonon branch temperatures, Waldecker $et~al$. developed a nonthermal lattice model for describing the microscopic energy flow in aluminium~\cite{PhysRevX.6.021003}. 
Their results show that using the two-temperature model to fit time-resolved experimental data may lead to misestimation of the electron-phonon coupling coefficient.
Drawing on the electron-phonon coupling two-temperature model, An $et~al$. extended it to study the phonon coupling thermal resistance between low-frequency out-of-plane phonons and high-frequency in-plane phonons in graphene~\cite{an2017a}.

Although above macroscopic models made great progress to explain non-equilibrium energy exchange and transport process between energy carriers in low-dimensional materials, they assumed the diffusive transport with infinite propagation speed in default~\cite{RevModPhysJoseph89}, which is questionable especially in nanosized materials or in low-temperature environments.
For example, when the characteristic size of the system is comparable to or much smaller than the phonon mean free path, ballistic transport dominates the heat transfer and non-Fourier heat conduction phenomena appear, such as the size effects~\cite{RevModPhys.90.041002,YANG201085,ZHANG2020,xu_length-dependent_2014,nomura_review}, graded thermal conductivity~\cite{yang_nanoscale_2015,ma_unexpected_2017,chuang2021graded} and thermal vortices~\cite{ZHANG20191366,shang_heat_2020}. 
When the characteristic response time of the system is comparable to or much smaller than the phonon relaxation time, the thermal wave phenomena occur~\cite{RevModPhysJoseph89}.

To realize precise simulations of the non-diffusive transport of energy carriers in materials, the Boltzmann transport equation (BTE) has become an effective tool~\cite{IJHMT_2006_EPcoupling_BTE,JHTelectron-phonon2009,PhysRevB.103.125412,PhysRevResearch.3.023072,JAP2016EP_review,PhysRevB.103.125412}.
Chen $et~al.$ developed a semiclassical two-step heating model to investigate thermal transport in metals caused by ultrashort
laser heating~\cite{IJHMT_2006_EPcoupling_BTE}. 
Instead of diffusive electron transport, three equations of the conservation of number density, momentum and energy are derived for the electron subsystem based on the electron BTE.
Tong $et~al.$ developed a numerical solution framework of the time-dependent BTE for modeling the ultrafast coupled dynamics of electrons and phonons~\cite{PhysRevResearch.3.023072}.
It was validated through simulations of pump-probe spectroscopy, x-ray diffuse scattering, and structural and phonon dynamics.
This method can accurately capture the complex phonon-phonon scattering and electron-phonon scattering processes, but it is computationally intensive.
Miao $et~al.$ found that there are over $20\%$ deviations in the electron-phonon coupling coefficients obtained through fitting the experimental data with the BTE and two-temperature model when the excitation pulse width is comparable to the relaxation time~\cite{MIAO2021121309,PhysRevB.103.125412}.
Based on the low-order Chapman-Enskog expansion of the BTE, Zhang $el~al.$ theoretically prove that only in the diffusive limit can the BTE recover the macroscopic two-temperature model, otherwise there must be some high-order time-space partial derivative terms~\cite{ZHANG2024123379}.
For example, the thermal wave phenomenon could appear when the system size is comparable to the mean free path which cannot be predicted by the two-temperature diffusive model.

Most of the previous studies based on the BTE focused on the ultra-fast laser heating in metal or semi-metal materials, ensuring the restoration of the typical two-temperature model at the macroscopic scale.
In addition, most studies focused on the lattice thermal conductivity of the materials as a whole, and less on the non-equilibrium and non-Fourier thermal conduction of each phonon branch, especially in nanosized or low-dimensional materials.
To this end, the non-equilibrium transport and phonon branch-resolved size effects are studied in this paper.
A multi-temperature kinetic model is developed. Taking single-layer suspended graphene as an example, the branch-resolved thermal conductivity and non-Fourier heat conduction are studied and discussed.

\section{Multi-temperature kinetic model}
\label{sec:BTEtheory}

A multi-temperature kinetic model is developed for electron-phonon coupling~\cite{ZHANG2024123379,MIAO2021121309,PhysRevB.103.125412,JAP2016EP_review},
\begin{align}
\frac{ \partial u_e }{\partial t} + \bm{v}_e \cdot \nabla u_e = \frac{ u_e^{eq}  -u_e }{\tau_e}  - \sum_{k=1}^{N} \frac{ G_{ep,k} }{2\pi} (T_e - T_{p,k}) + \frac{\dot{S} }{2\pi},  \label{eq:epBTE1} \\
\frac{ \partial u_{p,k} }{\partial t} + \bm{v}_{p,k} \cdot \nabla u_{p,k} = \frac{ u_{p,k}^{eq}  -u_{p,k} }{\tau_{p,k} } + \frac{ G_{ep,k} }{2\pi}  (T_e - T_{p,k})  ,
\label{eq:epBTE2}  
\end{align}
where the subscripts $e$ and $p$ represent electron and phonon, respectively.
Subscripts $k$ and $N$ are the index and total number of phonon branches, respectively. 
$u$ is the distribution function of energy density, $\bm{v}$ is the group velocity, $\tau$ is the relaxation time, $\dot{S}=\dot{S}(\bm{x},t)$ is the external heat source, $G_{ep}$ is the electron-phonon coupling constant.
$u_e^{eq}=  C_e T_e/(2 \pi) $ and $u_{p,k}^{eq}= C_{p,k} T_{lattice}/(2 \pi) $ are the equilibrium state, where $C$ is the specific heat, $T_{lattice}$ is the lattice temperature and $2\pi$ results from the sum of the solid angles in two-dimensional materials.
The phonon and electron scattering kernels both satisfy the energy conservation, 
\begin{align}
\int \frac{ u_{e}^{eq}  -u_{e} }{\tau_{e} } d\Omega &=0 ,  \\
\sum_{k=1}^{N}  \int \frac{ u_{p,k}^{eq} (T_{lattice})  -u_{p,k} }{\tau_{p,k} } d\Omega &=0 ,  \label{eq:phononscattering}
\end{align}
where $d\Omega$ represents the integral over the whole solid angle space.
Macroscopic variables, such as the electron temperature $T_e$, phonon branch-resolved temperature $T_{p,k}$, lattice temperature $T_{lattice}$ and electron heat flux $\bm{q}_e$, phonon branch-resolved heat flux $\bm{q}_{p,k}$, total heat flux $\bm{q}_{lattice}$, are obtained by taking the moment of distribution function~\cite{IJHMT_2006_EPcoupling_BTE,JHTelectron-phonon2009,PhysRevB.103.125412,PhysRevResearch.3.023072,JAP2016EP_review},
\begin{align}
T_e  = \frac{\int u_e  d\Omega }{C_e} ,  \quad   T_{p,k} = \frac{\int u_{p,k}  d\Omega }{C_{p,k}} ,  \quad 
T_{lattice} = \frac{ \sum_{k=1}^{N}  C_{p,k} T_{p,k} /  \tau_{p,k}   }{ \sum_{k=1}^{N} C_{p,k}/ \tau_{p,k}  } ,  \label{eq:latticeT}  \\
\bm{q}_e  = \int \bm{v}_e u_e  d\Omega   ,  \quad     \bm{q}_{p,k}  =  \int \bm{v}_{p,k} u_{p,k}  d\Omega   ,  \quad 
 \bm{q}_{lattice}   = \sum_{k=1}^{N}     \bm{q}_{p,k}  .  \label{eq:latticeheatflux}   
\end{align}
The lattice temperature is calculated based on the energy conservation of phonon scattering kernel~\eqref{eq:phononscattering}, which represents the energy exchange among various phonons under the framework of relaxation time approximation kinetic model.
{\color{black}{Note that the thermodynamic temperature cannot be well defined in non-equilibrium thermal systems.
The temperatures calculated by above formulas~\eqref{eq:latticeT} are more like the symbols of local energy density based on the local thermal equilibrium assumptions~\cite{Kubo1991statistical,ChenG05Oxford}.
Currently we are focusing on steady-state problems and systems with dimensions larger than $100$ nanometers.
The infinite time scale of the steady state and the spatial scale larger than $100$ nanometers mean that phonon-phonon scattering or energy exchange can still occur, thus defining a reasonable local equivalent equilibrium temperature~\cite{Kubo1991statistical,ChenG05Oxford}.

Although Fourier's law of thermal conduction may not necessarily hold true at the micro/nano scale, its formal diffusion solution is usually used to fit the detection signals in micro/nano scale thermal measurement experiments~\cite{chang_breakdown_2008,ZHANG2020,ziabari2018a,xu_length-dependent_2014,nomura_review}.
Ultimately, the overall temperature or the temperature of a specific phonon branch is derived. 
In other words, the temperature obtained from micro/nano scale thermal measurement experiments actually reflects the average local energy density at the heating spot or detection position.}}
Rigorously, the constitute relationship between the heat flux and temperature at the micro/nano scale should be used instead of diffusion equation~\cite{RevModPhysJoseph89,ChenG05Oxford}. 
Unfortunately, this constitute relationship is unknown at the micro/nano scale.
Hence the constitute relationship at the macroscopic scale is used to fit or define the physical variables at the micro/nano scale~\cite{nano_letters_2017_non_equilibrium,Distinguishing_AdvSci_2020,exp_photon_excitation2021,chang_breakdown_2008,ZHANG2020,ziabari2018a}, For example, the thermal conductivity along the radial direction in a nanosized homogeneous system is defined as the heat flux divided by the temperature gradient in the radial direction~\cite{yang_nanoscale_2015,ma_unexpected_2017,chuang2021graded},
\begin{align}
\kappa_e = \left|  \frac{q_e}{dT_e/dr}  \right|, \quad  \kappa_{p,k} = \left|  \frac{q_{p,k}}{d T_{p,k}/dr}  \right|,   \quad  
\kappa_{lattice} = \left|  \frac{q_{lattice}}{dT_{lattice}/dr}  \right|. \label{eq:localkappa}
\end{align}
An overall effective thermal conductivity along a particular direction at the micro/nano scale was also usually used~\cite{balandin_superior_2008,Distinguishing_AdvSci_2020,exp_photon_excitation2021,YANG201085,chang_breakdown_2008,ZHANG2020,PhysRevLett.87.215502,xu_length-dependent_2014}, that is, the product of the heat flux $q$ and the system size $L$ divided by the temperature difference $\Delta T$ between the two ends of the system along that direction,
\begin{align}
\kappa_{eff,e} = \left|  \frac{q_e L}{\Delta T} \right|, \quad  \kappa_{eff,p,k} = \left|  \frac{q_{p,k} L}{\Delta T} \right|,   \quad  
\kappa_{eff,lattice} = \left|  \frac{q_{lattice} L}{\Delta T} \right|.   \label{eq:effectivekappa}
\end{align}

\section{Results and discussions}
\label{sec:discussions}

Stationary heat conduction in a single-layer suspended graphene material at room temperature is studied under the multi-temperature kinetic model, which is solved numerically by a discrete unified gas kinetic scheme~\cite{GuoZl16DUGKS,ZHANG2024123379}.
More details of kinetic scheme and discretizations can be found in~\ref{sec:dugks}.
Thermophysical properties of electron and phonon in graphene materials at room temperature are shown in Table~\ref{parameters}, which are calculated by first-principle~\cite{PhysRevB.93.125432} and obtained from a previous paper~\cite{PhysRevB.98.134309}.
{\color{black}{A dimensionless analysis is also conducted to theoretically analyze the impact of input parameter on the heat conduction process in~\ref{sec:dimensionlessanalysis}.}}

\renewcommand{\arraystretch}{1.2}
\begin{table} 
\caption{Thermophysical properties of electron and phonon in graphene materials at room temperature.}
\centering
\begin{tabular}{|*{6}{c|}}
\hline
& $G_{ep}$ (W/(m$^3 \cdot$K))  &  $\lambda$ ($\mu$m)  & $C$ (J/(m$^3 \cdot$K)) & $\tau$ (ps)  &  $|\bm{v}|$ (m/s)   \\
\hline
LA &  1.000e14  & 0.802  & 1.900e5  & 7.080e1  & 1.133e4  \\
\hline
TA  & 1.000e12 &  0.192 &  3.200e5 &  2.470e1  & 0.776e4 \\
\hline
ZA  & 0.0   & 1.700 &  6.100e5 &  3.170e2  & 0.536e4 \\
\hline
LO &  6.000e14  & 0.082 &  3.000e4  & 1.000e1  & 0.817e4 \\
\hline
TO  & 2.700e15  & 0.110 &  2.000e4 &  1.200e1 &  0.913e4 \\
\hline
ZO  & 0.0 &  0.318 &  1.600e5 &  3.880e2  & 0.082e4  \\
\hline
Electron & / &  0.281 & 3.564e2 &  2.806e-1  & 1.0e6  \\
\hline  
\end{tabular}
\label{parameters}
\end{table}

\begin{figure}[htb]
\centering
\subfloat[]{\includegraphics[scale=0.095,clip=true]{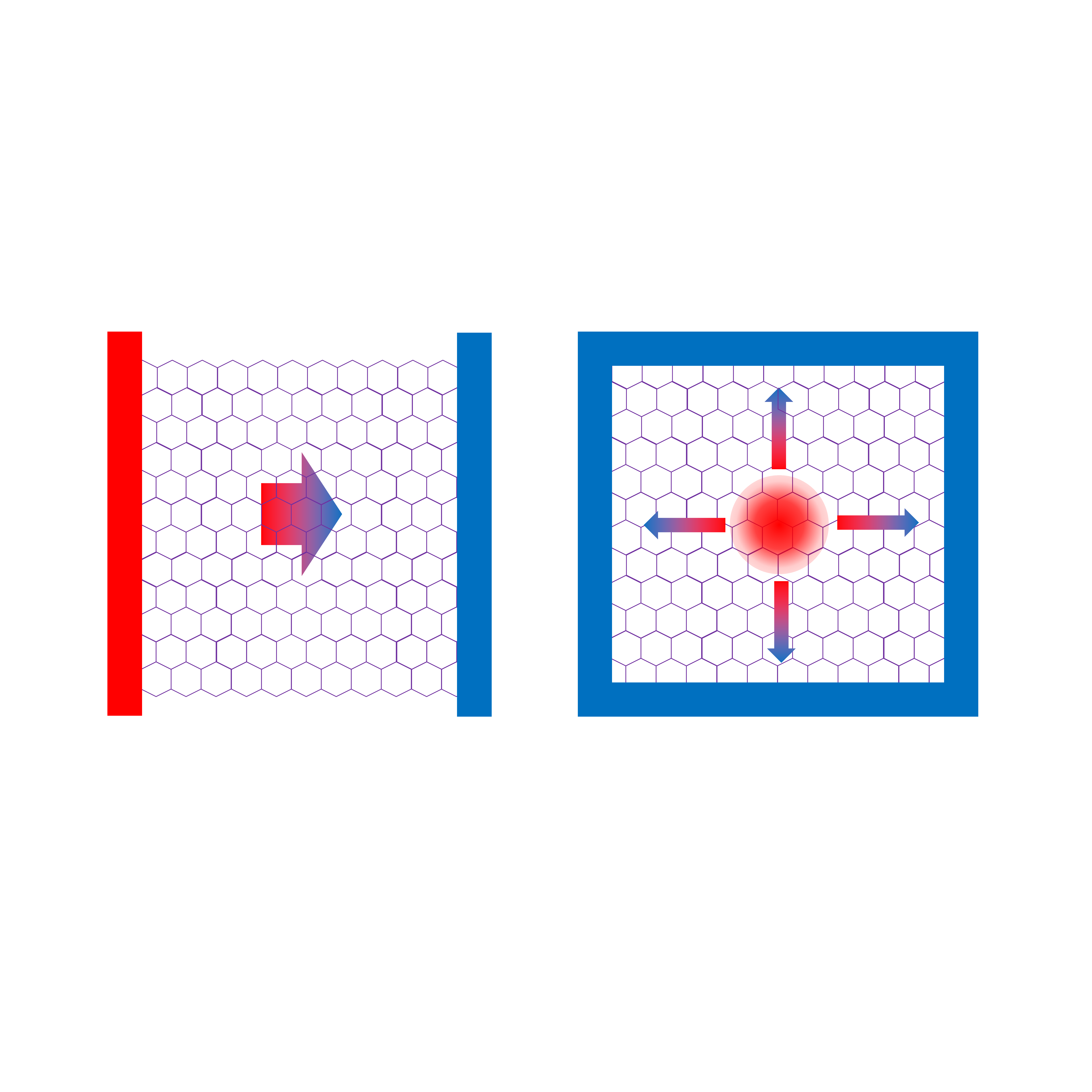}}~~~~~
\subfloat[]{\includegraphics[scale=0.095,clip=true]{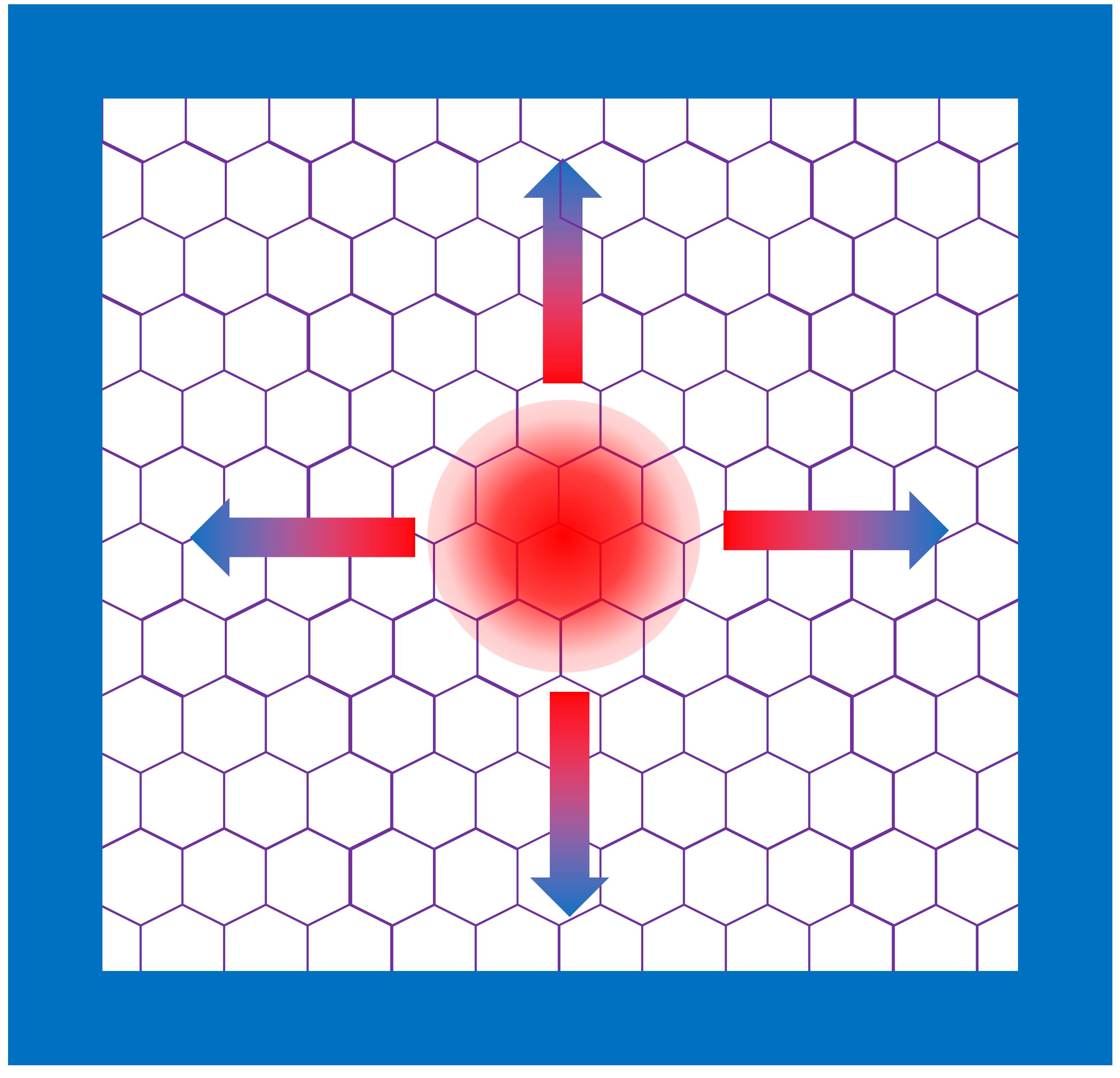}} 
\caption{(a) The two end sides of single-layer graphene materials are isothermal boundaries with fixed temperature $T_h=298$ K and $T_c=296$ K, respectively. (b) A Gaussian heating spot is added at the center and the four sides of single-layer graphene materials are isothermal boundaries with fixed environment temperature $297$ K.}
\label{graphenesettings}
\end{figure}

Firstly, thermal conduction in a thin film is simulated with different system sizes $L$, as shown in~\cref{graphenesettings}(a).
Initial temperature inside the domain is room temperature $T_0=297$ K.
The two end sides of graphene are thermalizing boundaries with temperature $T_h=298$ K and $T_c=296$ K, respectively.
Heat flows from the left to the right, just like the thermal bridge experiments~\cite{PhysRevLett.87.215502,xu_length-dependent_2014}.

\begin{figure}[htb]
\centering
\subfloat[100 $\mu$m]{\includegraphics[scale=0.30,clip=true]{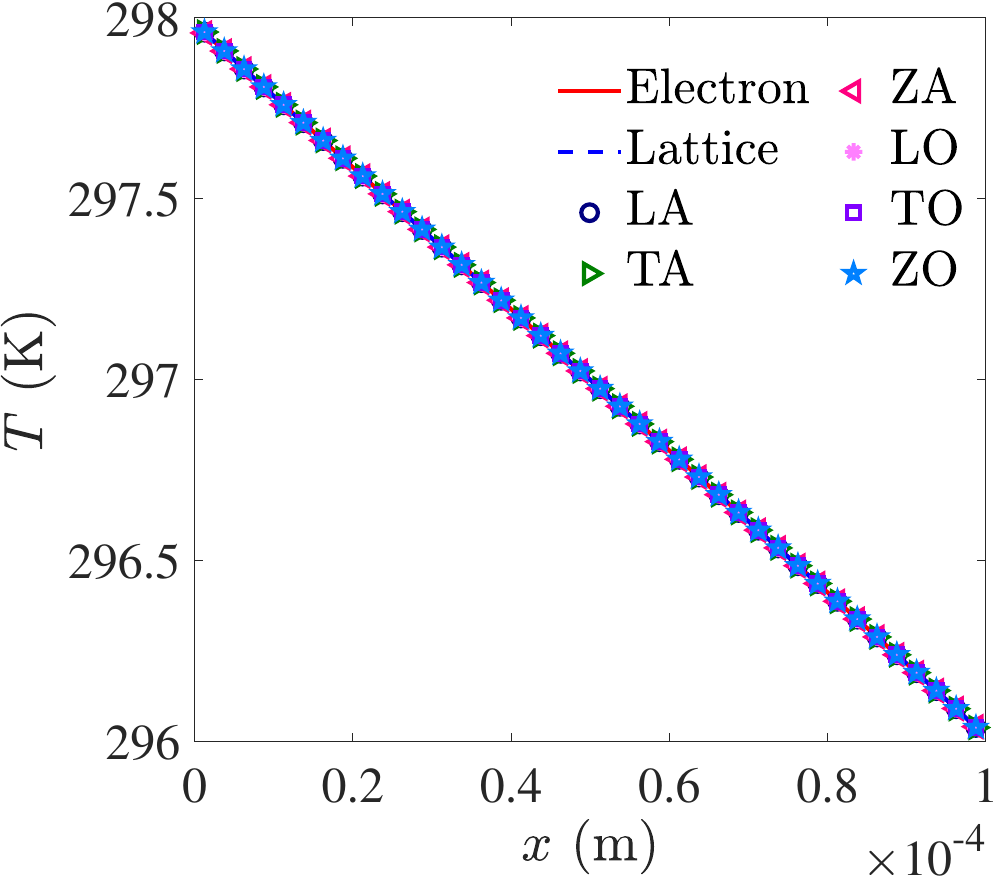}}~
\subfloat[$100$ nm]{\includegraphics[scale=0.30,clip=true]{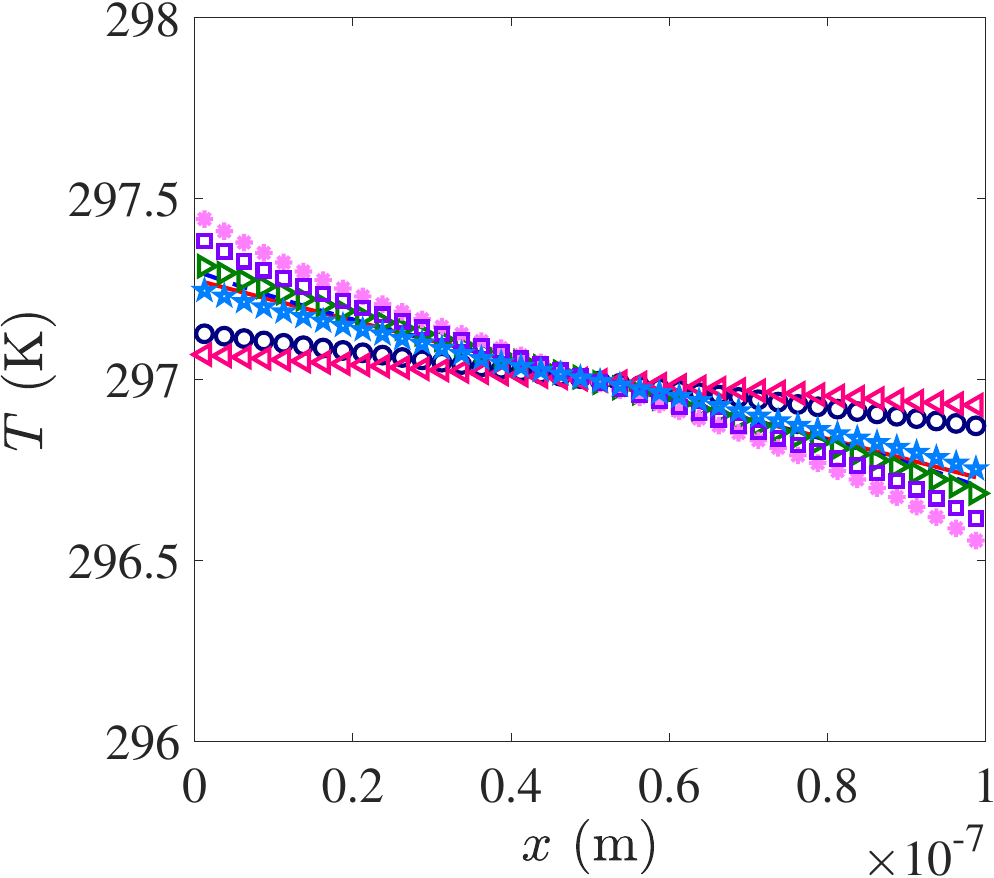}} \\
\subfloat[]{\includegraphics[scale=0.30,clip=true]{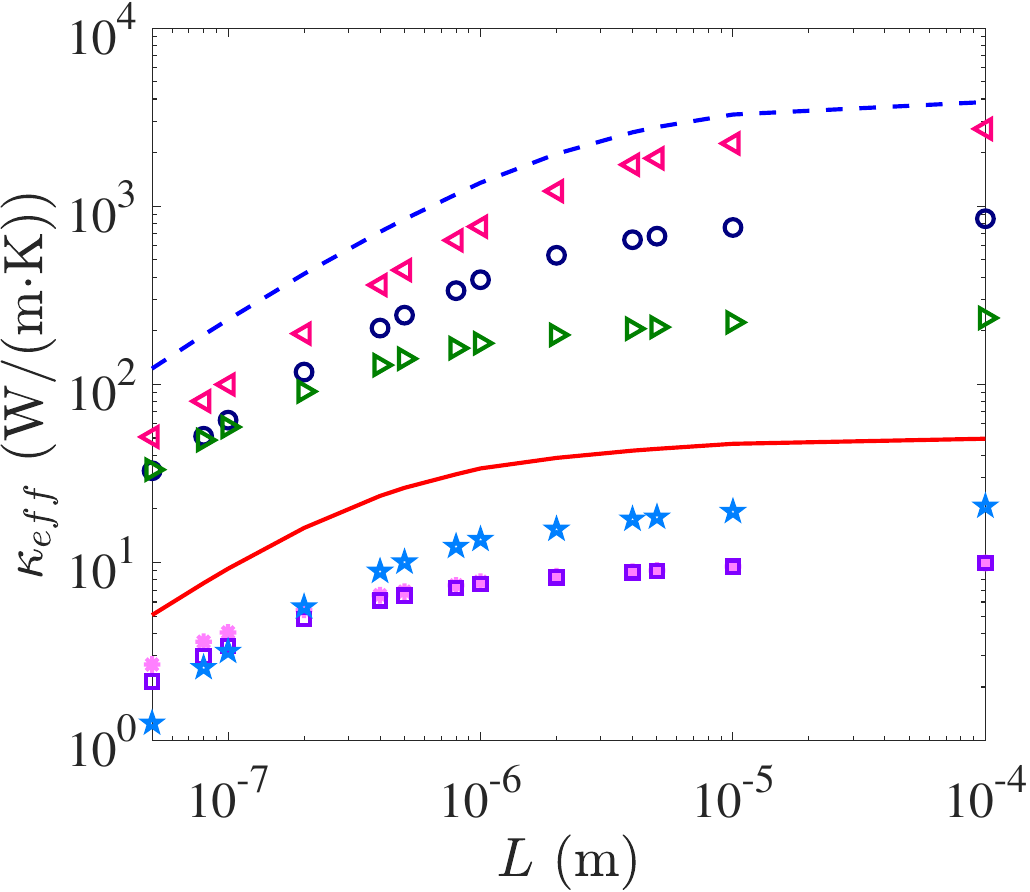}}~
\subfloat[]{\includegraphics[scale=0.30,clip=true]{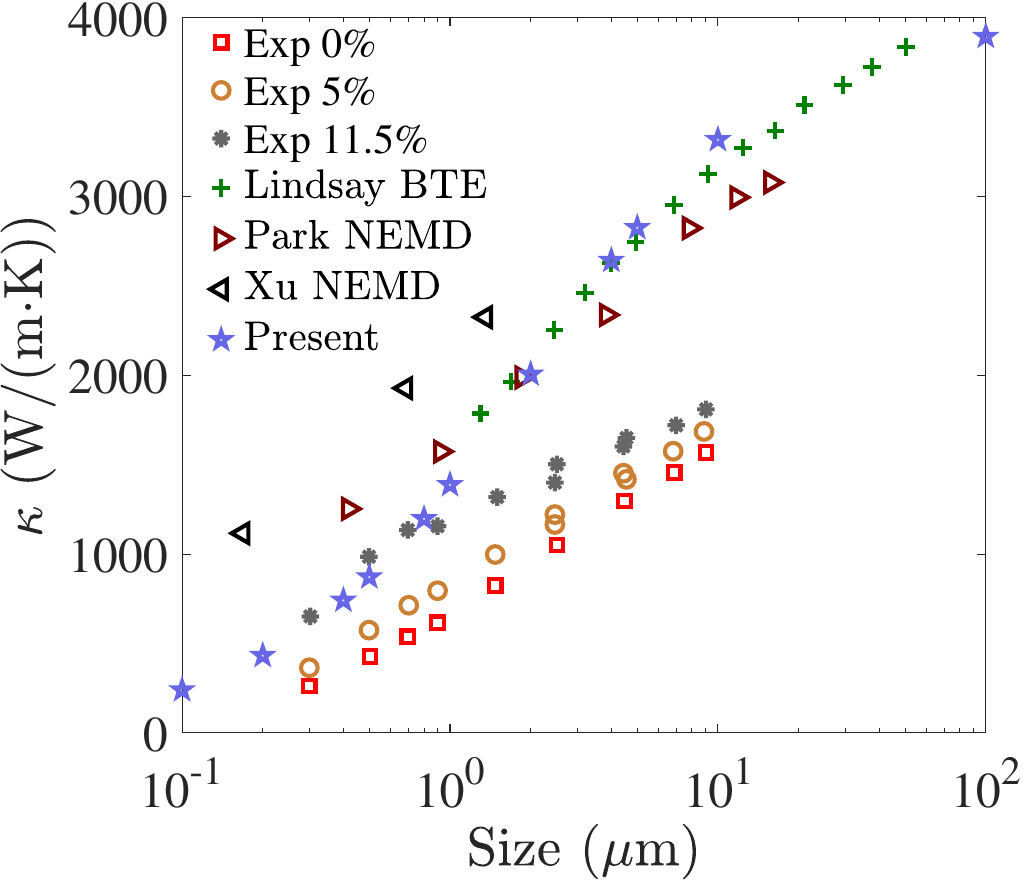}}~
\caption{Spatial distributions of (a,b) temperature and (c) branch-dependent effective thermal conductivity~\eqref{eq:effectivekappa}. (d) Size effects of single-layer suspended graphene at room temperature. Experimental data is obtained from Xu $et~al$~\cite{xu_length-dependent_2014} when assuming thermal contact resistance contributes $0\%$, $5\%$, $11.5\%$ to the total measured thermal resistance in a $9~\mu$m-long sample with fixed width $1.5~\mu$m. BTE and nonequilibrium molecular dynamics (NEMD) results are obtained from Refs.~\cite{PhysRevB.89.155426,NEMD_park2013,xu_length-dependent_2014}, respectively. `Present' is the total thermal conductivity of electron and phonon in this study.}
\label{temperaturesizes}
\end{figure}

The phonon branch-dependent and size-dependent thermal behaviors are simulated and discussed.
Figure~\ref{temperaturesizes}(a,b) shows the spatial distributions of temperature with different system sizes. 
When the system size is $100~\mu$m, the phonon-phonon, electron-electron, electron-phonon scattering processes are very frequent at this length scale.
Sufficient energy exchange among various particles leads to a thermal equilibrium between different phonon branches and electrons.
It can be found that the electron temperature, phonon branch temperature and lattice temperature are the same, and the numerical profiles keep linear without temperature slip near the boundaries, which is consistent with typical Fourier's law.
When the system size decreases from $100~\mu$m to $100$ nm, which is comparable to or smaller than the mean free paths of most phonon and electron, non-diffusive and non-equilibrium thermal effects happens.
On one hand, when particles are emitted from the boundary in the equilibrium state corresponding to the boundary temperature, they are almost not scattered with other particles within a spatial range of mean free path. 
This results in that the energy exchange among particles is insufficient and the boundary temperature cannot be efficiently transferred to the particles inside the geometry.
Consequently, a temperature slip appears near the boundaries.
The larger the mean free paths are, the larger the temperature slip is.
On the other hand, there are obvious temperature deviations between different phonon branches and electron due to the insufficient scattering processes.
In addition, the temperature distributions far from the boundaries keep almost linear regardless of phonon branches, electron or system sizes, which indicates that the thermal conductivity~\eqref{eq:localkappa} inside the domain is nearly a constant.

Figure~\ref{temperaturesizes}(c) shows the size- and branch-dependent effective thermal conductivity~\eqref{eq:effectivekappa}.
When the system size is $100~\mu$m, the effective thermal conductivities converge regardless of phonons and electrons.
When the system size decreases, being comparable to or smaller than mean free paths, the free transport process of particles is blocked by the geometry boundary in advance.
Namely, the effective mean free paths are shortened by the boundary scattering, so that it can be found that the effective thermal conductivity decreases when the system size decreases from tens of microns to tens of nanometers regardless of electron or phonon or each phonon branch.
When the system size is slightly larger than $1~\mu$m, the effective thermal conductivity almost reaches convergence for TA and optical phonons because their mean free paths are smaller than those of LA and ZA.
Compared to the optical phonons, the effective thermal conductivity of acoustic phonons are larger.
ZA phonon branch contributes most to the thermal conduction regardless of system size due to the larger mean free path and specific heat.

{\color{black}{Figure~\ref{temperaturesizes}(d) shows the size-dependent effective thermal conductivity of single-layer suspended graphene at room temperature obtained by BTE~\cite{PhysRevB.89.155426}, nonequilibrium molecular dynamics~\cite{NEMD_park2013,xu_length-dependent_2014} or experimental measurements~\cite{xu_length-dependent_2014}.
It can be found that the present results are in good agreement with Lindsay's results predicted by phonon BTE with input parameters calculated by first-principle~\cite{PhysRevB.89.155426} when the system size is in the range of $1~\mu$m to $100~\mu$m.
All theoretical calculation results are higher than the thermal bridge experimental measurement results~\cite{xu_length-dependent_2014} when the system size is in the range of $0.3~\mu$m to $9~\mu$m. 
This is due to several factors: 1) the presence of contact thermal resistance and impurity doping in the experiment will reduce the thermal conductivity to a certain extent, but the theoretical simulations are more concerned with undoped materials; 2) the width is infinite in the present simulation, but the width of the experimental sample is only $1.5$ microns~\cite{xu_length-dependent_2014}. When the length is much larger than the width, the smaller width may inhibit phonon transport and thus reduce thermal conductivity; 3) different teams have measured different thermal conductivities of graphene~\cite{graphene_review_conduction}, and there are measurement errors and sample differences in the experiment. 
For example, Balandin $et~al$ conducted Raman experiments and the results showed that the thermal conductivity of suspended single-layer graphene at room temperature is between $3000$ and $5000$ W/(m$\cdot$K)~\cite{balandin_superior_2008}.
Generally speaking, the predicted values in this paper are in the range of previous experimental data and the non-equilibrium molecular dynamics simulation results.}}

\begin{figure}[htb]
\centering
\subfloat[$40~\mu$m]{\includegraphics[scale=0.300,clip=true]{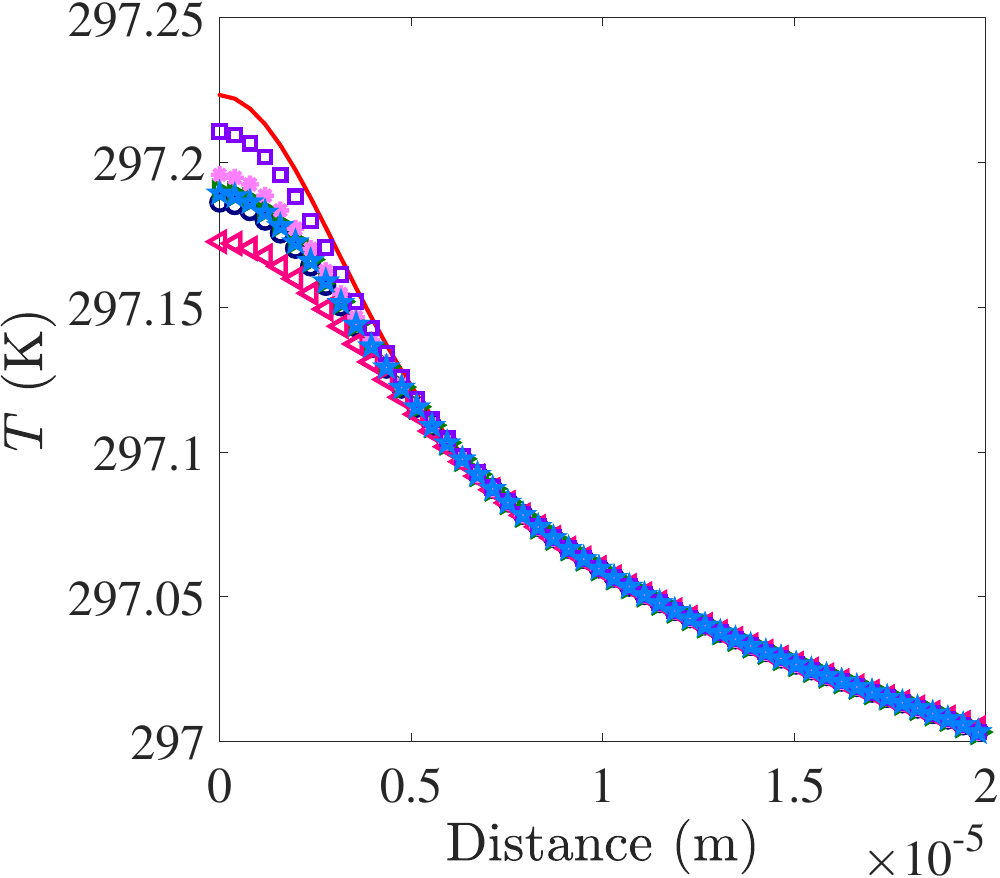}}~
\subfloat[$4~\mu$m]{\includegraphics[scale=0.300,clip=true]{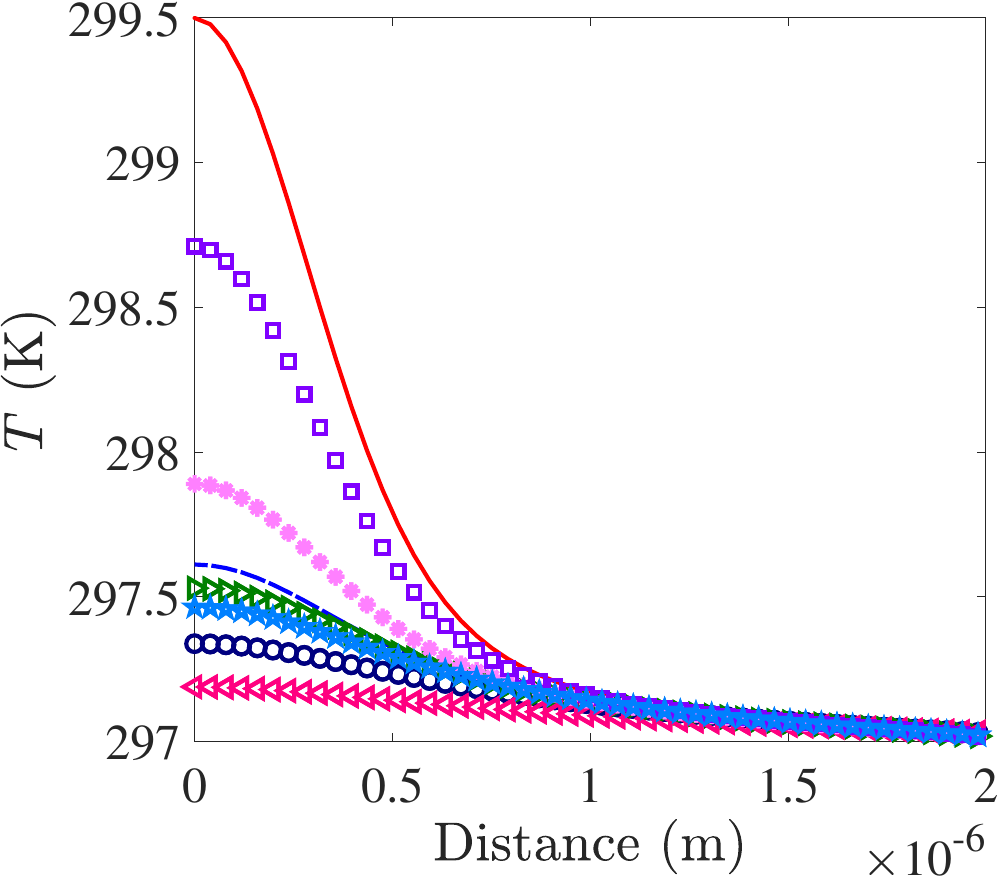}}~
\subfloat[$0.4~\mu$m]{\includegraphics[scale=0.300,clip=true]{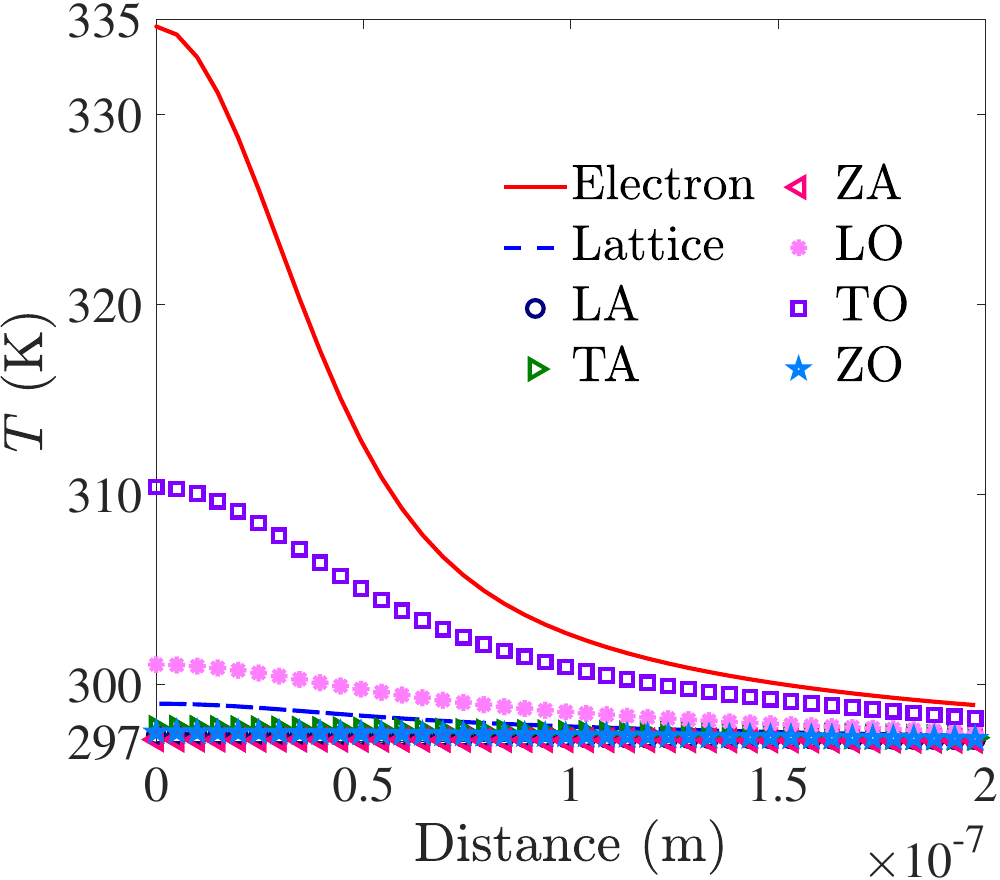}}\\
\subfloat[$40~\mu$m]{\includegraphics[scale=0.300,clip=true]{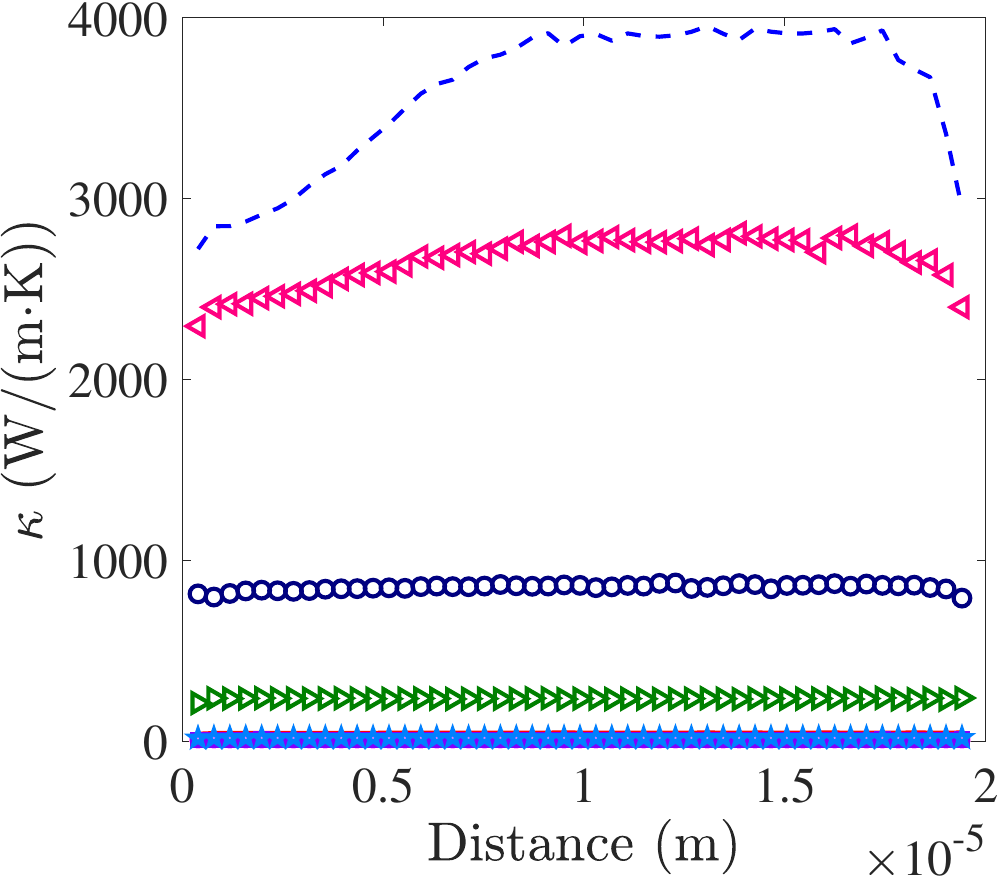}}~
\subfloat[$4~\mu$m]{\includegraphics[scale=0.300,clip=true]{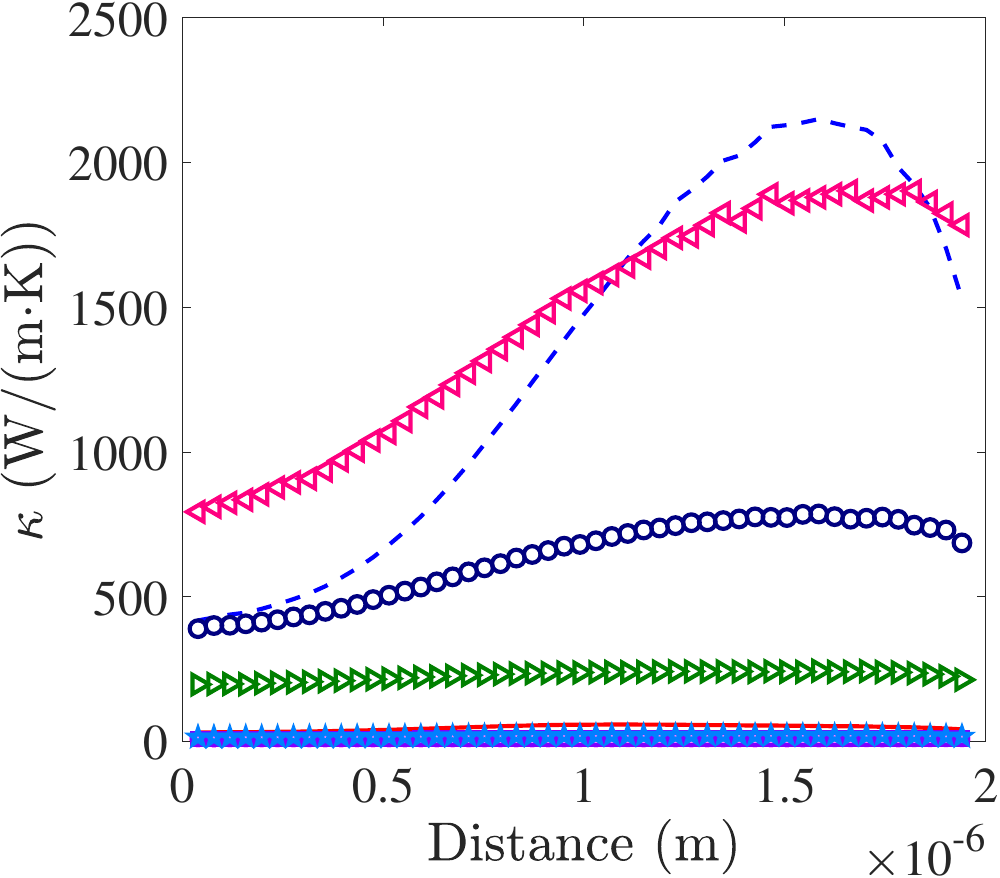}}~
\subfloat[$0.4~\mu$m]{\includegraphics[scale=0.300,clip=true]{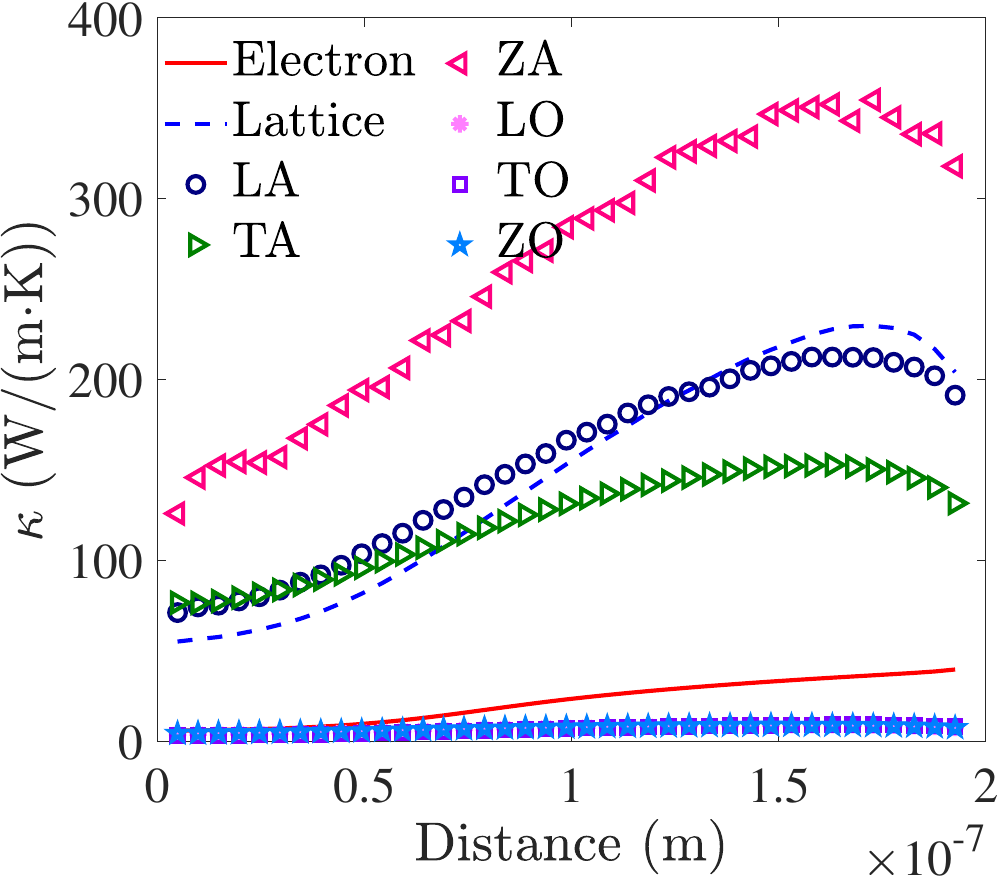}}~
\caption{Spatial distributions of (a-c) temperature and (d-f) thermal conductivity~\eqref{eq:localkappa} along the radial direction with different system sizes, where the horizontal axis is the distance from the center of geometry shown in~\cref{graphenesettings}(b).}
\label{diskkappasizes}
\end{figure}

Secondly, motivated by the Raman experiments~\cite{balandin_superior_2008,Distinguishing_AdvSci_2020,exp_photon_excitation2021}, a continuous Gaussian heating pump pulse is implemented on the single-layer suspended graphene,
\begin{align}
\dot{S} = \frac{1250}{ r_{pump}^2 } \times \exp \left( - \frac{2  r^2 }{r_{pump}^2 }  \right) .
\end{align}
where $r$ is the distance from the center, $r_{pump}=L/8$ is the heating pump radius, $L$ is the side length of systems, as shown in \cref{graphenesettings}(b).
The four end sides are all thermalizing boundaries with room temperature $297$ K.

The non-Fourier heat conduction or phonon branch-resolved size effects are simulated and discussed.
As reported in the last example (\cref{temperaturesizes}c), the smaller the system size is, the smaller the effective thermal conductivity is.
Consequently, when the system size decreases, the poorer heat dissipations efficiency leads to a higher temperature rise under the same power input, as shown in~\cref{diskkappasizes}(a-c).
Similarly, the deviations between electron temperature, lattice temperature and phonon branch temperature increases significantly when the system size decreases due to insufficient scattering or low energy exchange frequency.

The thermal behaviors in planar (\cref{graphenesettings}a) or radial ~(\cref{graphenesettings}b) homogeneous system are quite different when the system size is comparable to or smaller than mean free path.
It can be observed from~\cref{diskkappasizes}(a-c) that for a given nanosized radial system, the temperature deviations between various particles gradually decreases from the inner to outer.
The thermal conductivity~\eqref{eq:localkappa} along the radial direction increases from the inner to the outer.
Namely, the graded thermal conductivity appears in a homogeneous nanomaterials with fixed size~\cite{yang_nanoscale_2015,ma_unexpected_2017,chuang2021graded}, which exactly breaks the typical Fourier's law.
This graded thermal conductivity phenomenon does not appear in planar geometry.
In other words, different thermal measurement techniques, for example thermal conducts from the left-end heat source to the right-end heat sink~\cite{PhysRevLett.87.215502,xu_length-dependent_2014}, or from the central heat source to the surrounding heat sinks~\cite{balandin_superior_2008,Distinguishing_AdvSci_2020,exp_photon_excitation2021}, may change the non-Fourier heat conduction characteristics in nanosized materials and affect the thermal conductivity values obtained by fitting the experimental signals within different models~\cite{PhysRevB.100.085203,MIAO2021121309,ziabari2018a}.

We analyze above anomalous thermal phenomena from the perspective of insufficient phonon scattering or ballistic transport. 
For planar geometry~(\cref{graphenesettings}a), in the direction perpendicular to the temperature gradient, the system size remains unchanged and the particle transport and scattering mechanisms are almost the same. 
Along the temperature gradient direction, the particles emitted from the high-temperature heat source will definitely be absorbed by the low-temperature heat sink in the ballistic limit, and the particles emitted from the low-temperature heat sink will definitely be absorbed by the high-temperature heat source, too.
However, in the radial homogeneous system~(\cref{graphenesettings}b), on one hand, in the direction perpendicular to the temperature gradient, that is, the tangent direction, the circumference of the tangential circle gradually increases from the inside to the outside. 
As the distance from the center increases, the geometric space in which particles can fly freely increases, and the mean free paths of particles are no longer greatly suppressed by the boundary scattering, so the heat conduction efficiency increases and the thermal conductivity gradually increases from the inside to the outside.
On the other hand, along the temperature gradient direction, in the ballistic limit, the particles emitted from the high-temperature heat source will definitely be absorbed by the low-temperature heat sink, but due to the unequal inner and outer sizes, the particles emitted from the low-temperature heat sink may directly return to the low-temperature heat sink without passing through the high-temperature heat source area. 
Note that these particles do not contribute to the local heat flux, but contribute to the local temperature.
This unequal size between the inside and outside affects the particles transport and scattering in the ballistic regime and also changes the heat conduction characteristics.
{\color{black}{The graded thermal conductivity indicates that the key issue in dissipating heat from nanoscale hotspots in chips is to address the low thermal conductivity near the heat source~\cite{chen1996}. 
For example, liquid cooling microchannels are directly embedded inside the chip or high thermal conductivity materials are used as the substrate for the transistors~\cite{van2020co,HUA2023chapter}.}}

Furthermore, it can be found that the thermal conductivity of ZA branch is even higher than the lattice thermal conductivity when the system size is $4~\mu$m or $0.4~\mu$m, as shown in~\cref{diskkappasizes}(e,f).
For ZA branch, the phonon mean free paths are much larger than this system size so that they suffer ballistic transport process with little phonon scattering thermal resistance.
A huge temperature slip near the heat source areas appears due to the ballistic transport, which indicates that a smaller temperature gradient inside the domain, as shown in~\cref{diskkappasizes}(c).
Compared to the other phonon branches, smaller scattering thermal resistance, larger specific heat and longer mean free path of ZA branch lead to a larger thermal conductivity. 
For lattice heat conduction as a whole, six phonon branches contribute to the heat flux so that the lattice heat flux is larger than that of each branch (Eq.~\eqref{eq:latticeheatflux}).
However, the mean free path of optical phonons are very small, comparable to or larger than the system size, so that they suffer more phonon scattering process which leads to larger thermal resistance.
Optical phonons contribute a little to the heat flux, but sufficient scattering makes the temperature between various phonon branch go to a constant.
In other words, the lattice temperature gradient inside the domain becomes larger than that of ZA branch under the optical phonon scattering.
Hence, it is possible to predict that the thermal conductivity of a single phonon branch is larger than the lattice thermal conductivity as a whole within certain system sizes.
This result also suggests that it may be possible to make heat conduction more efficient by regulating the heat source to selectively excite specific phonon modes~\cite{APLnonthermal2020,wan2024manipulating}.

\section{Conclusion}
\label{sec:conclusion}

Non-equilibrium phonon transport and branch-resolved size effects in single-layer graphene materials are studied.
A multi-temperature kinetic model is developed for capturing the branch-dependent electron-phonon coupling.
Results show that the present kinetic model could describe the branch-resolved phonon transport process from tens of nanometers to tens of microns.
Thermal behaviors of each phonon branch are different in nanosized graphene, including temperature slips near the boundaries and the size-dependent effective thermal conductivity.
Phonon ZA branch contributes highest to the thermal conduction regardless of the system size.
In addition, for a nanosized homogeneous graphene with a hotspot at the center, the branch-dependent thermal conductivity increases from the inside to the outside even if the system size is fixed.
When the system size is hundreds of nanometers, the thermal conductivity of ZA branch is even higher than the lattice thermal conductivity.

\section*{Author Statements}

\textbf{Chuang Zhang}: Supervision, Conceptualization, Investigation, Methodology, Numerical analysis, Funding acquisition, Writing - original draft.
\textbf{Houssem Rezgui}: Numerical analysis, Writing-review \& editing.
\textbf{Meng Lian}: Methodology, Writing-review \& editing.
\textbf{Hong Liang}: Conceptualization, Numerical analysis, Funding acquisition, Writing-review \& editing.

\section*{Conflict of interest}

No conflict of interest declared.

\section*{Acknowledgments}

C. Z. acknowledges the support of National Natural Science Foundation of China (No. 52506078).
H. L. acknowledges the support of National Natural Science Foundation of China (No. 12572285).
The authors acknowledge Beijng PARATERA Tech CO.,Ltd. for providing HPC resources that have contributed to the research results reported within this paper.
C. Z. acknowledges the members of online WeChat Group: Device Simulation Happy Exchange Group, for the communications on BTE simulations.

\appendix

\section{Discrete unified gas kinetic scheme}
\label{sec:dugks}

To numerically solve the multi-temperature kinetic model, a discrete unified gas kinetic scheme~\cite{guo_progress_DUGKS,GuoZl16DUGKS} is introduced, where the solid angle, time and spatial position spaces are discretized into a lot of small pieces under the framework of finite volume method.
Taking an integral of the kinetic equation over a control volume $i$ from time $t_m$ to $t_{m+1}=t_{m}+ \Delta t$, and the mid-point rule is used for the time integration of flux term and the trapezoidal rule is used for other in order to achieve second-order temporal accuracy.
Consequently, the discreteized kinetic equations are written as follows,  
\begin{align}
&\tilde{u}_e^{m+1} =  \left(  u_{e,i,n}^{m+1}  -\frac{\Delta t}{2}   H_{i,n}^{m+1}   \right) =
 -\frac{\Delta t}{V_i} \sum_{j \in N(i)} \left(  \bm{v}_e \cdot \mathbf{n}_{ij} u_{e,ij,n}^{m+1/2} S_{ij} \right) + \left( u_{e,i,n}^{m} +\frac{\Delta t}{2}   H_{i,n}^{m}  \right),   \label{eq:dpBTE1} \\
&\tilde{u}_{p,k}^{m+1} =  \left( u_{p,k,i,n}^{m+1} -\frac{\Delta t}{2}  F_{i,k,n}^{m+1} \right) =
- \frac{\Delta t}{V_i} \sum_{j \in N(i)} \left(  \bm{v}_{p,k} \cdot \mathbf{n}_{ij} u_{p,k,ij,n}^{m+1/2} S_{ij} \right)  + \left( u_{p,k,i,n}^{m}  + \frac{\Delta t}{2}  F_{i,k,n}^{m}  \right), \label{eq:dpBTE2} 
\end{align}
where $H= (u_e^{eq}  -u_e )/ \tau_e - \sum_{k=1}^{N}  G_{ep,k}(T_e - T_{p,k})/ (2 \pi)  + \dot{S} / (2 \pi) $, $F_k=( u_{p,k}^{eq}  -u_{p,k} ) /  \tau_{p,k} +  G_{ep,k} (T_e - T_{p,k})/ (2 \pi)$, $n$ represents the index of discretized solid angle space, $V_i$ is the volume of the cell $i$, $N(i)$ denotes the sets of neighbor cells of cell $i$, $ij$ denotes the interface between cell $i$ and cell $j$, $S_{ij}$ is the area of the interface $ij$, $\mathbf{n}_{ij}$ is the normal unit vector of the interface $ij$ directing from cell $i$ to cell $j$, $\Delta t$ is the time step and $m$ is an index of time step.

In order to obtain the distribution function at the cell interface at the mid-point time step, taking an integral of the kinetic equation from time $t_m$ to $t_{m+1/2}=t_{m}+ \Delta t/2$ along the characteristic line with the end point $\bm{x}_{ij}$ locating at the center of the cell interface $ij$ between cell $i$ and cell $j$,
\begin{align}
&\bar{u}_e^{m+1/2} = u_{e}^{m+1/2} (\bm{x}_{ij} ) -\Delta t/4   H^{m+1/2}(\bm{x}_{ij} )= u_{e}^{m} (\bm{x}_{ij} -\bm{v}_e \Delta t/2 ) + \Delta t/4   H^{m }(\bm{x}_{ij} -\bm{v}_e \Delta t/2 )  , \label{eq:BTEfaces1}   \\
&\bar{u}_{p,k}^{m+1/2} = u_{p,k}^{m+1/2} (\bm{x}_{ij} )- \Delta t/4 F_k^{m+1/2}(\bm{x}_{ij} ) = u_{p,k}^{m} (\bm{x}_{ij} -\bm{v}_{p,k} \Delta t/2 )  + \Delta t/4   F_k^{m }(\bm{x}_{ij} -\bm{v}_{p,k} \Delta t/2 ) . \label{eq:BTEfaces2}
\end{align}
The right hand side of Eqs.(\ref{eq:BTEfaces1},\ref{eq:BTEfaces2}) at the $m-$time step can be directly obtained by numerical interpolations, such as the van Leer or upwind scheme, least square method.
Taking an integral of Eqs.(\ref{eq:BTEfaces1},\ref{eq:BTEfaces2}) over the whole solid angle space leads to
\begin{align}
&\frac{ \Delta t}{4}  \dot{S} + \int  \bar{u}_e   d\Omega   =   \left( C_e +\frac{\Delta t}{4}  \sum_{k=1}^{N} G_{ep,k}  \right) T_e  - \frac{\Delta t}{4} \sum_{k=1}^{N} \left( G_{ep,k} T_{p,k} \right) ,  \label{eq:momentface1}  \\
&\int    \bar{u}_{p,k} d\Omega  =  - \frac{\Delta t}{4}  G_{ep,k} T_e  + \left(\frac{\Delta t}{4} \frac{C_{p,k}}{\tau_{p,k}}  +C_{p,k} +  \frac{\Delta t}{4}  G_{ep,k}   \right)  T_{p,k}  -\frac{\Delta t}{4} \frac{C_{p,k}}{\tau_{p,k}} T_{lattice}, \label{eq:momentface2}
\end{align}
where the left hand sides of the above two equations are already known.
Combining these two equations and Eq.(3), the electron temperature $T_e$, phonon branch temperature $T_{p,k}$ and lattice temperature $T_{lattice}$ at the cell interface at the mid-point time step can be calculated by iteration method.

After the equilibrium states or macroscopic variables at the cell interface are obtained, $u_e^{m+1/2}$ and $u_{p,k}^{m+1/2}$ at the cell interface can be updated directly.
Then $\tilde{u}_{e}^{m+1}$ and $\tilde{u}_{p,k}^{m+1}$ at the cell center can be calculated based on Eqs.~(\ref{eq:dpBTE1},\ref{eq:dpBTE2}).
Taking an integral of Eqs.~(\ref{eq:dpBTE1},\ref{eq:dpBTE2}) over the whole solid angle space leads to
\begin{align}
&\frac{ \Delta t}{2} \dot{S} + \int  \tilde{u}_e   d\Omega   =   \left( C_e +\frac{\Delta t}{2}  \sum_{k=1}^{N} G_{ep,k}  \right) T_e  - \frac{\Delta t}{2} \sum_{k=1}^{N} \left( G_{ep,k} T_{p,k} \right) ,  \label{eq:momentcenter1} \\
&\int    \tilde{u}_{p,k} d\Omega  =  - \frac{\Delta t}{2}  G_{ep,k} T_e  + \left(\frac{\Delta t}{2} \frac{C_{p,k}}{\tau_{p,k}}  +C_{p,k} +  \frac{\Delta t}{2}  G_{ep,k}   \right)  T_{p,k}  -\frac{\Delta t}{2} \frac{C_{p,k}}{\tau_{p,k}} T_{lattice}, \label{eq:momentcenter2} 
\end{align}
where the left hand side of the above two equations are already known.
Combining the these two equations and Eq.(3), the electron temperature $T_e$, phonon branch temperature $T_{p,k}$ and lattice temperature $T_{lattice}$ at the cell center at the next time step can be calculated by iteration method.
Then $u_e^{m+1}$ and $u_p^{m+1}$ can be updated.

Compared to our previous paper of DUGKS for electron-phonon coupling~\cite{ZHANG2024123379}, the biggest improvement of the present work is the introduction of phonon branch-resolved thermal properties.
A bigger matrix containing electron and all phonon branches information has to be invoked and solved iteratively~\cite{zhang_discrete_2019} when calculating the macroscopic fields from the distribution function, namely, Eqs.(\ref{eq:momentface1} ,\ref{eq:momentface2}) and Eqs.(\ref{eq:momentcenter1} ,\ref{eq:momentcenter2}).

In the simulations of Fig.1(a), $10-40$ uniform cells are used to discrete the spatial domain and $80-16$ directions are used to equally discrete the two-dimensional solid angles when system size increases from $10$ nm to $100~\mu$m.
In the simulations of Fig.2(b), $101^2$ uniform cells are used to discrete the spatial domain and $16-80$ directions are used to equally discrete the two-dimensional solid angles.
Time step is $\Delta t= 0.50 \times \Delta x/|\bm{v}_e| $, where $\Delta x$ is the minimum cell size.
The grid independence test shows that the current numerical discretizations can meet the computational accuracy requirements. 

\section{Dimensionless analysis}
\label{sec:dimensionlessanalysis}

There is no analytical solutions of BTE for most problems, and it is difficult to directly determine the change pattern of temperature or heat flux with time and spatial position based on input parameters, especially for complex irregular geometric structures or materials. 
This paper discusses the influence of input parameters on the phonon thermal conduction characteristics based on the principle of dimensional analysis. 

Make a dimensionless analysis of the multi-temperature BTE model without external heat source,
\begin{align}
\frac{ \partial u_e^* }{\partial t^*} + |\bm{v}_e^*| \bm{s} \cdot \nabla_{\bm{x}^*} u_e^* &= \frac{ u_e^{eq,*}  -u_e^* }{\tau_e^* } - \sum_{k=1}^{N} \frac{ (T_e^* -T_{p,k}^* )/(2 \pi) }{1/G_{ep,k}^*  } ,  
\label{eq:dimensionalessbte1}   \\
\frac{ \partial u_{p,k}^* }{\partial t^*} + |\bm{v}_{p,k}^*| \bm{s}  \cdot \nabla_{\bm{x}^*} u_{p,k}^* &= \frac{ u_{p,k}^{eq,*}  -u_{p,k}^* }{\tau_{p,k}^* } +  \frac{ (T_e^* -T_{p,k}^* )/(2 \pi) }{1/G_{ep,k}^*  },  \label{eq:dimensionalessbte2}
\end{align}
where $\bm{s}$ is the unit directional vector and the thermal physical parameters of electrons are regarded as the reference variables $v_{\text{ref} } =|\bm{v}_e|$,  $C_{\text{ref} }= C_e$, so that
\begin{align}
t^* &= \frac{t}{t_{\text{ref} }},  \quad& G_{ep,k}^* &= \frac{ t_{\text{ref} }  }{ C_{\text{ref} }/G_{ep,k}  } , \quad&  \bm{x}^* &= \frac{ \bm{x}  }{ L_{\text{ref} }},   \\
u_{p,k}^*&= \frac{u_{p,k} }{ C_{\text{ref} } T_{\text{ref} }  },  \quad&  \tau_{p,k}^* &= \frac{ \tau_{p,k}  }{ t_{\text{ref} }}, \quad&  \bm{v}_{p,k}^* &= \frac{ \bm{v}_{p,k}  }{ v_{\text{ref} }},  \\
u_e^*&= \frac{u_e}{ C_{\text{ref} } T_{\text{ref} }  },  \quad&  \tau_e^* &= \frac{ \tau_e  }{ t_{\text{ref} }}, \quad&  \bm{v}_e^* &= \frac{ \bm{v}_e  }{ v_{\text{ref} }},  \\
T_{p,k}^* &= \frac{T_{p,k} }{T_{\text{ref}} } ,  \quad&  T_e^* &=\frac{T_e}{T_{\text{ref} }}, \quad &  t_{\text{ref} }&= \frac{ L_{\text{ref} } }{v_{\text{ref}} },
\end{align}
where $L_{\text{ref} }$ is the system characteristic size and $T_{\text{ref}}$ is the reference temperature.

{\color{black}{Based on the dimensionless analysis of the kinetic model, the transient evolution processes of the system are determined by these dimensionless variables, i.e., $\tau_e^*$, $|\bm{v}_e^*|$, $\tau_{p,k}^*$, $|\bm{v}_{p,k}^*|$, $1/G_{ep,k}^*$. 
For steady-state problem, the final convergent solutions are totally determined by $|\bm{v}_e^*|\tau_e^*$, $|\bm{v}_{p,k}^*|\tau_{p,k}^*$,  $|\bm{v}_e^*| /G_{ep,k}^*$, $|\bm{v}_{p,k}^*| /G_{ep,k}^*$. 
When $1 \gg \tau_e^*$ and $1 \gg  |\bm{v}_e^*| \tau_e^*$, diffusive electron transport happens.
When $1 \gg \tau_{p,k}^*$ and $1 \gg  |\bm{v}_{p,k}^*| \tau_{p,k}^*$ is satisfied for single phonon branch, diffusive phonon transport happens for that branch. 
When this condition is satisfied for all phonon branches, the branch-dependent phonon temperature tends to a constant, but may be different from electron temperature.
When $G_{ep,k}^* \gg 1$, $1 \gg |\bm{v}_e^*|/G_{ep,k}^*$ and $1 \gg |\bm{v}_{p,k}^*|/G_{ep,k}^*$ is satisfied for single phonon branch, the deviations between electron and phonon temperature of that branch tend to zero.
When this condition is satisfied for all phonon branches, the branch-dependent phonon temperature tends to a constant, and be equal to electron temperature.}}

When both electron and phonon suffer diffusive transport processes, the distribution function can be approximated as according to the first-order Chapman-Enskpg expansion
\begin{align}
u_e^*  &\approx  u_e^{eq,*} - \tau_e^* \left( \frac{ \partial u_e^{eq,*} }{\partial t^*} + |\bm{v}_e^*| \bm{s} \cdot \nabla_{\bm{x}^*} u_e^{eq,*}  +  \sum_{k=1}^{N}  \frac{ (T_e^* -T_{p,k}^* )/(2 \pi) }{1/G_{ep,k}^*  }   \right) , \label{eq:btece1}  \\
u_{p,k}^*  &\approx  u_{p,k}^{eq,*} - \tau_{p,k}^* \left( \frac{ \partial u_{p,k}^{eq,*} }{\partial t^*} + |\bm{v}_{p,k}^*| \bm{s} \cdot \nabla_{\bm{x}^*} u_{p,k}^{eq,*}  - \frac{ (T_e^* -T_{p,k}^* )/(2 \pi) }{1/G_{ep,k}^*  }   \right) . \label{eq:btece2}
\end{align}
Combined above six equations and taking an integral of the BTE (\ref{eq:dimensionalessbte1},\ref{eq:dimensionalessbte2}) over the whole first Brillouin zone, we can get
\begin{align}
\frac{ \partial  }{\partial t^* } \left< u_e^{eq,*} - \tau_e^* \left( \frac{ \partial u_e^{eq,*} }{\partial t^*}  +  \sum_{k=1}^{N}  \frac{ (T_e^* -T_{p,k}^* )/(2 \pi) }{1/G_{ep,k}^*  }    \right)   \right> - \nabla_{\bm{x}^*} \cdot \left< |\bm{v}_e^*||\bm{v}_e^*| \bm{s} \bm{s} \tau_e^* u_e^{eq,*} \right>   =-   \sum_{k=1}^{N}  \frac{ T_e^* -T_{p,k}^*   }{1/G_{ep,k}^*  }   , \\
\frac{ \partial  }{\partial t^* } \left< u_{p,k}^{eq,*} - \tau_{p,k}^* \left( \frac{ \partial u_{p,k}^{eq,*} }{\partial t^*}  -\frac{ (T_e^* -T_{p,k}^* )/(2 \pi) }{1/G_{ep,k}^*  }   \right)   \right> - \nabla_{\bm{x}^*} \cdot \left< |\bm{v}_{p,k}^*||\bm{v}_{p,k}^*| \bm{s} \bm{s} \tau_{p,k}^* u_{p,k}^{eq,*} \right>   =  \frac{  T_e^* -T_{p,k}^*   }{1/G_{ep,k}^*  }  ,
\end{align}
where $<>$ represents the integral over the whole first Brillouin zone.
Then the associated macroscopic heat conduction is
\begin{align}
C_e \frac{ \partial T_e  }{\partial t }  - \tau_e  \frac{ \partial^2 U_e  }{\partial t^2 } - \tau_e \frac{ \partial  }{\partial t }  \left(\sum_{k=1}^{N} G_{ep,k}  (T_e -T_{p,k})  \right)    &= \nabla \cdot ( \kappa_e \nabla T_e )  -  \sum_{k=1}^{N} G_{ep,k}  (T_e -T_{p,k})   , \label{eq:TTMce1}    \\
C_{p,k} \frac{ \partial T_{p,k}  }{\partial t }  - \tau_{p,k}  \frac{ \partial^2 U_{p,k}  }{\partial t^2 } + \tau_{p,k} \frac{ \partial  }{\partial t } \left( G_{ep,k}  (T_e -T_{p,k})  \right)     &= \nabla \cdot ( \kappa_{p,k} \nabla T_{p,k} )  + G_{ep,k} (T_e -T_{p,k} ), \label{eq:TTMce2}
\end{align}
where $\kappa = \int C |\bm{v}|^2 \tau /2  d\bm{K}$ is the bulk thermal conductivity.
When $\sqrt{\tau_e^*} \ll 1$, $\sqrt{\tau_{p,k}^*} \ll 1 $, $\tau_e^* G_{ep,k}^*  \ll 1$, $\tau_{p,k}^* G_{ep,k}^*  \ll 1$ for all phonon branches and electron, the first order term of relaxation time in Eqs.~(\ref{eq:TTMce1},\ref{eq:TTMce2}) could be removed and the multi-temperature kinetic model can recover the macroscopic multi-temperature model in the diffusive limit~\cite{ZHANG2024123379},
\begin{align}
C_e \frac{ \partial T_e  }{\partial t }  &= \nabla \cdot ( \kappa_e \nabla T_e )  -\sum_{k=1}^{N} G_{ep,k}  (T_e -T_{p,k})      \\
C_{p,k} \frac{ \partial T_{p,k}  }{\partial t }  &= \nabla \cdot ( \kappa_{p,k} \nabla T_{p,k} )  + G_{ep,k} (T_e -T_{p,k} ) .
\end{align}

\bibliographystyle{elsarticle-num-names_clear}
\bibliography{phonon}
\end{document}